# Development of a novel matrix-based methodology for system engineering: A case study

Hossein Sabzian[1]

Seyyed Mostafa Seyyed Hashemi[2]

Ehsan Kamrani[3]

**Abstract:**

**Developing a structured method for analyzing various aspects of a system requires a novel methodology. This study is aimed at developing such as methodology through combining two major matrix methods, namely, Design Structure Matrix (DSM) and Interface Structure Matrix (ISM). Through this paper, a business process modeling method is applied to turn a real work project to a process model. Then that process model is written in two various matrix forms of DSM and ISM. These two matrices are analyzed by two types of algorithm for extracting activity levels and sub-processes. In the end, a Mixed Matrix Model (MMM) is built upon these activity levels and sub-processes, which can be used as a framework for the engineering of real-world systems.**

[1] - Department of Progress Engineering, Iran University of Science and Technology, Tehran, Iran (corresponding author: hossein_sabzian@pgre.iust.ac.ir)
[2] - Tadbir Economic Development Group, Tehran, Iran
[3] - Wellman Center for photomedicine Harvard-MIT Health Sciences and Technology, Harvard Medical School, Harvard University, Cambridge, MA, USA

# 1- Introduction

Accurate identification of systems and enhancement of their efficiency have always been considered the most important concern of managers at every organizational level. Complexity is one of the outstanding features of current socio-economic systems. To understand what systems are and how they work, researchers have developed several methods. Some researchers prefer network analysis among these methods as the main representatives of this methodology, program evaluation, and review technique (PERT) and critical path method (CPM) have a high ability in the analysis of projects. But their primary weakness is the inability to cover the projects with the backward flow (Wiest & Levy, 1977). Efforts made to eliminate shortcomings of PERT/CRM methods provided the basis for the creation of graphical evaluation and review technique (GERT). The method has a structure similar to PERT/CPM methods, but with the difference that it can display and analyze backward flows. Use of this method has a great convenience, but the multiplicity of documents complicates the information analysis. Therefore, the difficulty of analyzing information is considered the main disadvantage of GERT methodology. Parallel to these developments, matrix methods have been introduced gradually so that convenience, accuracy, and efficiency of these methods have attracted a wide range of analysts. Matrix methods involve a range of square matrices such as design structure matrix (DSM), domain mapping matrix (DMM), and multi-domain matrix (MDM) as well as non-square matrix such as interface structure matrix (ISM) (Kreimeyer, Eichinger, & Lindemann, 2007).

Tall structures, information flow complexity, large volumes of paperwork, the plurality of departments and units, the multiplicity of teams and individuals, and sometimes overlapping and unclear duties and activities of the employees have made the analysis of socio-economic systems very difficult. Some organizations, such as telecommunication companies are in such a situation. Senior managers, middle or even low ranking ones of such entities need to identify the correct working flows to decide on the units or departments under his/her responsibility.

These points raise a question of *"is it possible to design a model capable of presenting upstream and downstream activities as well as their constituent sub-processes clearly by covering various aspects of the system and help managers to systematically analyze it regarding those aspects?"*

## 2- History of DSM

The concept of DSM was introduced in the 60s and was studied more in the 80s by Stewart (Steward, 1981). DSM was not seriously taken into consideration until 1990[1].

Nowadays, it has been proved, DSM is one of the most enduring and outstanding achievements in engineering design. DSM has been the subject of several studies. In 2001, Browning (2001) examined applications of different DSMs (Browning, 2001). Chen *et al.* (2003) used the concept of DSM to manage new product development projects. So, scheduling related tasks were dramatically improved and the impacts of deferred tasks were analyzed with greater accuracy (C.-H. Chen, Ling, & Chen, 2003) A group from National Aeronautics and Space Administration (NASA) and MIT University, in a study conducted at NASA, used DSM to investigate the Pathfinder Spacecraft (Brady, 2002). Bilalis *et al.* (2004) used the concept of DSM to display characteristics associated with innovation resulting from the product, process, and approach of the management (Bilalis, Maravelakis, Antoniadis, & Moustakis, 2004). Cho and Eppinger (2005) presented a process model for the analysis of design projects. This model was based on DSM and considered the stochastic nature and resources constraints (Cho & Eppinger, 2005).

---

[1] The reintroduction of DSM was largely due to the growth of design theory studies.

Chen and Liu (2005) developed a numerical DSM that could quantify dependencies and their coupling strength (K.-M. Chen & Liu, 2005). Karniel *et al*. (2005) used DSM to break down a complex engineering problem (3D surface fitting) into sub-problems. These minor issues were solved optimally in a manner that was created by DSM (Karniel, Belsky, & Reich, 2005). Pektaş and Pultar (2006) used the parameter-based DSM in construction projects for a better understanding of the design process(Pektaş & Pultar, 2006). Kreimeyer *et al*. (2007) investigated numerous applications of matrix methods and DSMs (Kreimeyer et al., 2007). Another group used DSM for the analysis of single-purpose cameras. Finding conventional approaches across a product family was their primary objective in this study (Alizon, Moon, Shooter, & Simpson, 2007). *Sosa et al.* (2007) used DSM for the analysis of commercial aircraft jet engine in the study conducted at a jet engine manufacturer company in America (Sosa, Eppinger, & Rowles, 2007). Suh (2007) used DSM to analyze the digital printing system in a study conducted at the Xerox Corporation in America. Recognizing the impact of new technology on the current architecture of the product was taken into account as the main objective of their study (Suh, De Weck, & Chang, 2007). *Schmidt et al*. (2009) used DSM to analyze the school construction subject in a study conducted at BSF Company in Britain. The increase in understanding of system architecture to lead design requirements and identify alternatives as well as matching the design to potential future changes were considered two main goals of their study (Schmidt III, Austin, & Brown, 2009)

3- **What of DSM**

DSM is a square matrix of n*n in which n represents the number of entries in the system such as components of a product, tasks, and activities at a project, and teams in an organization. Each entry of the matrix structure represents a specific type of relationship between two elements (e.g., components of a product, etc.) in the system. For example, two pieces belong to the same assembly line make an assembled connection or two activity of an information flow create information flow dependence. Displaying complex dependencies by DSM allows the system to be divided into manageable sub-systems. System analysts can gain a lot of valuable information through the analysis of communication within and among sub-systems. According to the functionality, DSMs can be divided into four component-, team-, activity-, and parameter-based type, whereas they are classified as fixed DSM and time-based DSM according to the time. This classification is presented in Figure 1.

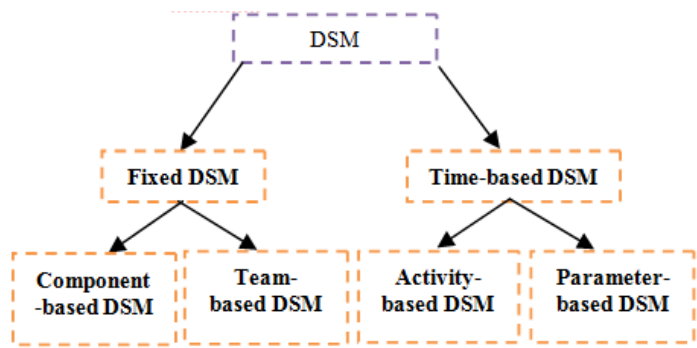

Figure 1: Classification of DSM [16]

Component-based DSM models the system architecture based on components or sub-systems and their interactions. Team-based DSM models organizational structures according to human resources or groups and their interactions. Activity-based DSM models processes and network activities based on activities and information flow and other dependencies. Low-level communication among designs and parameters decisions, equations, and interchange of semi-routine parameters can be modeled using the parameter-based DSM. Fixed DSM indicates interactions of system elements simultaneously, such as components of a product structure or parts of an organization. In the time-based DSM, rows and columns represent the flow over time. Upstream activities of a process have priority over downstream activities. When referring to communications, terms such as "feed forward" and "feedback" will be used to show the direction. Creating a DSM requires a lot of information regarding the system under study. Attaining such information about products and processes that include thousands of components or activities, if possible, will be very difficult and time-taking. Extraction of interactions among a system's elements is usually derived from two sources.

Interchangeability of matrix and graph enables researchers to quickly write process models in the matrix form. A process model is a directed network that displays information flows within the network as one of the subsets of the structured analysis and design technique (SADT)[1] to derive process models, **Integrated Definition for Process Description Capture Method** (IDEF3)[2] method has had wide applications in business process modeling. The notation system of IDEF3 allows system experts to show input/output streams and control processes in the network.

    Due to analyzing the interactions among elements of a system, DSM only provides specific information and just covers some certain aspects of the system. To overcome this shortcoming, it is necessary to use other matrix methods. Interface structure matrix (ISM), as a matrix that can analyze dissimilar elements of a system, can be used as a suitable supplement. Unlike DSM that displays the interactions of component-component (and activity-activity), ISM shows the interactions of component- interface (or activity-interface). Interactions of component-interface (or activity-interface) illustrate a different view of the dependency. ISM's information processing (i.e., clustering) leads to clusters similar to those produced by DSM.

In many cases, obtaining relevant information through interfaces or creating an ISM is far easier than a DSM. Also, the interpretation of a large DSM is difficult, while a large ISM can be easily analyzed (Jose

---

[1] SADT is a type of system engineering methodology which models the system in a functions hierarchy using activity models and data models. For a comprehensive study, see (Marca & McGowan, 1987; Ross, 1977)
[2] For a detailed study, see (Mayer et al., 1995)

& Tollenaere, 2005). Chen and Liu (2005) studied the effect of interfaces on product development strategies. When product structure interfaces[1] are determined, designing of parallel modules is made possible. This issue dramatically reduces the time of the design process cycle.

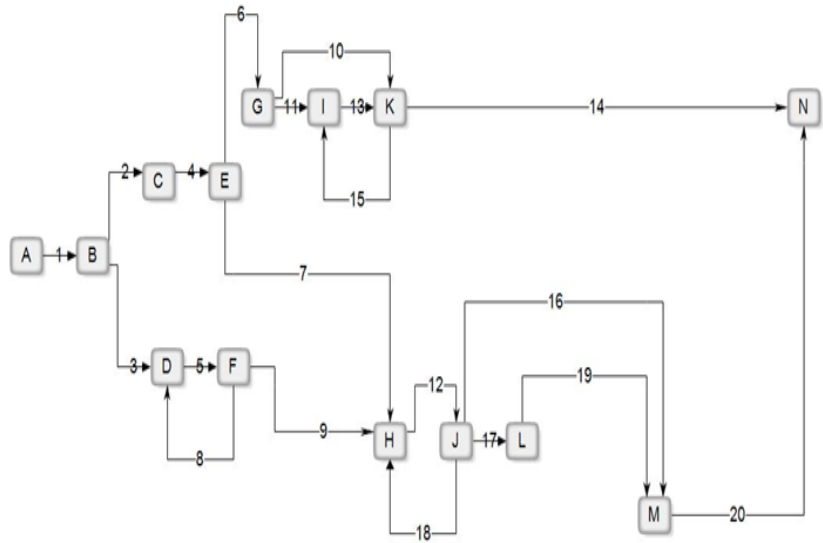

Figure 2: A process model with 14 activities (A to N)

## 4- A Case Study

### 4-1- Extraction of the Process Model of a Real Project through IDEF3

A process model with 14 activities (A to N) has is drawn in Figure 2, which has been developed by interviewing with experts of a

---

[1] They include input-output and output-/control communication flows.

*Telecommunication company*[1] and direct observation of the workflows. As can been seen in Figure 2, activities and dependencies are respectively shown by rectangles and arrows used in the IDEF3 Standard. According to the DSM pattern, a special dependence such as arrow 5 can be considered as the output of activity *D* and the input of activity *F*. It should be noted that the IDEF3 notation system has been used for the model of arrow 3 in which information (dependency) flows from left to right and up to down arrows indicate control flows. IDEF3 method can display input/output and control flows. In contrast to PERT/CPM, the issue has great practicality in real-world modeling processes. Control flows are often mistaken as input/output information flows and thus lead to unnecessary delays in the activity. The process model of Figure 2 can be analyzed by ISM and DSM.

## 4-2- Writing the Process Model into DSM

Identification of activity levels and communication between upstream and downstream activities provides valuable information for managers. The quality of such a communication process significantly influences the overall time of the process execution. In addition to the decomposition of a process model into the different activity levels, researchers can divide a process model into smaller processes (sub-processes). Extracting sub-processes of a process model prepares useful information about *the scheduling of the process model*, the role and *importance of its sub-processes* and therefore provides a useful model for an appropriate allocation of resources.

By writing down a process model in the form of DSM and sorting it through Triangulation Algorithm (TA), activity levels *(L1 ... LN)* of a process model would be extracted. By writing down a process model in ISM form and clustering it through Cluster Determination Algorithm (CDA), sub-processes *(S1 ... SM)* can be identified. Since activities are embedded in sub-processes, it is possible to simultaneously show activities and sub-processes of a process model. Such a simultaneous presentation of activities and sub-processes of an organization makes its systematic analysis easier.

---

[1] The name of company has not been presented to protect its anonymity.

The interactions among input and outputs of activities (A to N) of the process model of *Figure 1* are presented in *Figure 2* by DSM.[1]

Figure 3: **DSM of the process model of** Figure 2

The matrix shown in Figure *3* is written by the IR/FAD (Inputs into Row/Feedback above Diagonal) method. In the matrix, outputs and inputs, respectively, are indexed in columns and rows, and feedback flows are displayed above the original diameter. For example, if the entry of *(f, d)* is equal to 1, then the output of column *d* is the input of raw *f*. So *f* is dependent on *d*.

---

[1] It should be noted that output/control relationships have not been considered in the matrix because the time scale of their occurrence is different with time of input to output processing.

## 4-3- Extraction of Activity Levels by Sorting DSM Through Triangulation Matrix

There are a few algorithms for sorting and setting DSMs. Topological sorting algorithm (TSA) and triangulation algorithm (TA) are two algorithms for setting DSMs (Kusiak, 2008)). TSA has a relatively higher speed, but this algorithm can only be used for acyclic directed graph[1]. So, it cannot be used to sort the process model of *Figure 2*. In addition to the appropriate speed and ease of use, TA allows users to search circulations (cycles), and sort circular directed graphs (i.e., those graphs with a feedback circulation or dependence) efficiently (Kusiak, 2008). Meanwhile, all activity levels of a process model can be extracted and classified by TA.

### 4-3-1- Triangulation Algorithm

TA is more applicable than TSA because it can also be used for graphs with circulation. The algorithm is useful to analyze the structure of process models and determine activity levels both in downstream and in upstream(Kreimeyer et al., 2007). To present the algorithm, knowing several terms such as original activity (OA) and destination activity (DA) are necessary. An activity is called OA if there is no activity before that. OAs can easily be identified in incidence matrices. If the $i^{th}$ row of the incidence matrix has only one non-empty entry (diagonal entry), so $i$ is an OA. Activity is known as a DA if there is no activity after that. DAs can simply be recognized in incidence matrices[2], too. If the $j^{th}$ column of the incidence matrix has only one non-empty entry (diagonal entry), so $j$ is a DA. If there is no OA in the matrix, there will be at least one cycle or iteration (Kusiak, 1999).

---

[1] An acyclic directed graph is a graph which does not have any circulating flow.
[2] An incidence matrix is a matrix that represents the relationship between two classes of objects.

### 4-3-2- Implementation Stages of Triangulation Algorithm

This algorithm enables users to find cycles of an incidence matrix and triangulate it simultaneously. The algorithm is presented in a way that is easy to implement.

Step 0: Start with the current order of activities (1, 2..., n);
Step 1: If all activities are underlined, stop the algorithm. Identify the OA or DA. If OA or DA is not found, go to step 5;
Step 2: Use the Sorting Rule for the activity identified in step 1;
Step 3: underline the activity identified in step 1;
Step 4: remove the row and column related to the activity which had been underlined from the incidence matrix (see step 1) and go back to step 1;
Step 5: Find a cycle (or an iteration);
Step 6: Integrate all cycle's activities in a single activity;
Step 7: Integrate corresponding rows and columns in the incidence matrix and go back to step 1;
Step 8: Level activities and cycles according to the precedence relationship.

### 4-3-3- Sorting Rule

If an activity is an OA, shift it to the far left side in the existing order of underlined activities. If an activity is a DA, shift it to the far right side in the current order of underlined activities.

After doing the algorithm, identify and draw activity levels through 3 steps below.

1. Draw a circle (or a rectangular) around every single activity or each cycle;
2. Identify activities of each similar level (Do not create any relationship between them);
3. Title (name) the activities of each similar level;

Now the matrix of *Figure 3* can be clustered by TA as the matrix of *Figure 4* and the diagonal block pattern can be shown (Kusiak & Wang, 1993)

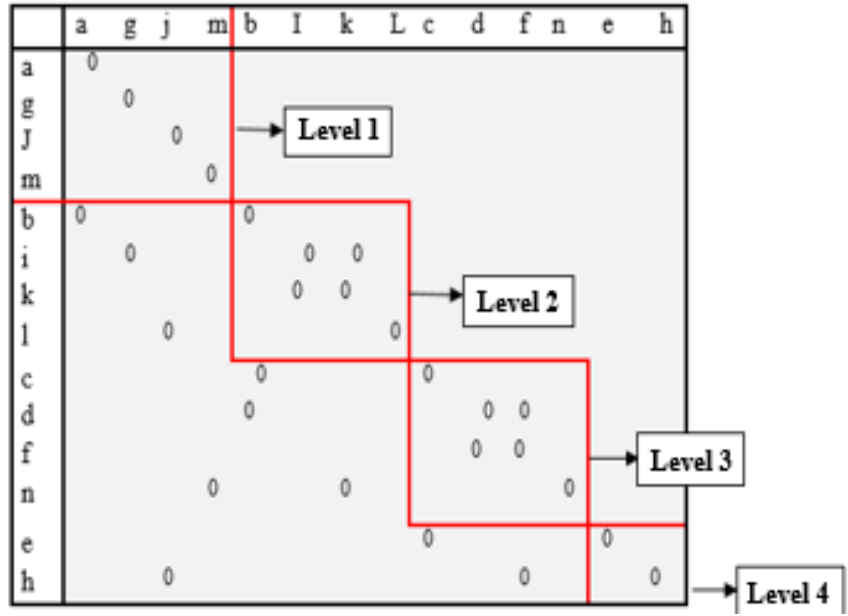

Figure 4: Sorting DSM of Figure 3 through TA

Each cluster represents the level of activity. They have been shown with four different colors in *Figure 4*. Also, interactions between clusters have been demonstrated by stars outside of painted boxes. All activities have been depicted as Fig. 5.

|  |  |  |  |
|---|---|---|---|
| A | B | C | E |
| G | I | D | H |
| J | K | F |  |
| M | L | N |  |
| ( Level 1 ) | ( Level 2 ) | ( Level 3 ) | ( Level 4 ) |

Figure 5: Activity levels of the process model of Figure 2

### 4-4- Writing the Process Model into ISM

ISM is a matrix made up of two components, including the interface or relationship and the object or entity. The interface or relationship refers to nature or relationship type of an object with another object. Unlike DSM, this matrix is usually not square. ISM has the characteristic that the element (entry) of an ISM is usually obtainable from databases and can be extracted through interviews with experts and authorities in each part of the project or the system.

To better explain ISM, it is assumed that the process of *Figure 2* has been completely indexed in Figure 7. The simplest way to create the matrix of Fig. 6 is respectively, the allocation of marks *I*, *O*, and *C* to input, output, and control for all interfaces associated with each activity.

For instance, *I*, *O*, and *C* are assigned to, respectively, interfaces 9, 12, and 18 for activity *h*. This means that the activity *h* has control, input, output, and input relations with interfaces 7, 9, 12, and 18, respectively. To facilitate comparison to DSM, Fig. 6 has been reduced to Figure 7, and only input/output interfaces have been indexed.

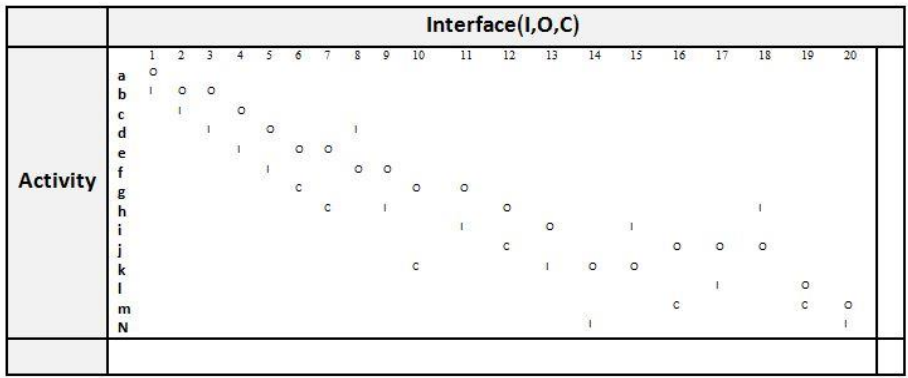

*Figure 6: ISM related to the process model of Figure 2*

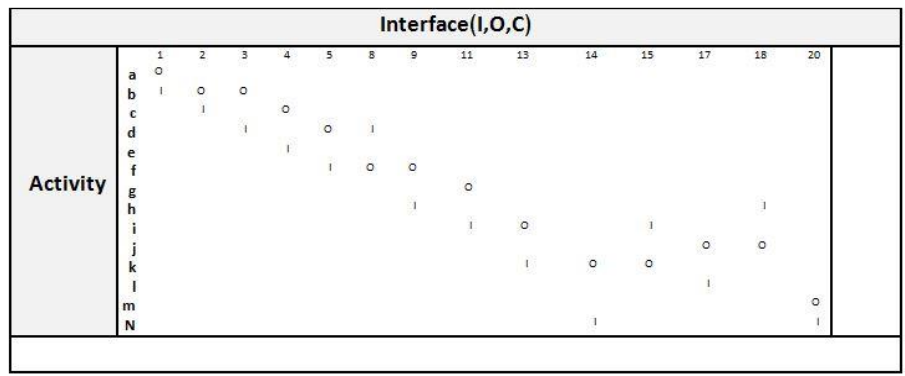

*Figure 7: Reduced ISM of Figure 6*

### 4-4-1- Extraction of Sub-Processes by Clustering ISM through CDA

Contrary to DSM that there are a few methods to cluster it, ISM model can be solved with different soft wares. There are many algorithms to cluster ISM that among them, it can be pointed to CDA. This algorithm was introduced by Kusiak and Cho in 1987. The modified form of the algorithm is effective to cluster matrices that do not have a diagonal structure (Kusiak, 1987). Clustering process models through the algorithm lead to extraction and classification of sub-processes.

### 4-4-2- Implementation Stages of CDA

Step 1: Show the number of iterations with ($k = i$).
Step 2: Select the row $i$ of matrix $A(k)$ and draw a horizontal line ($hi$) on it.
Step 3: Draw a vertical line ($vj$) on each marked entry (non-zero) adjoining the horizontal line ($hi$).
Step 4: Continue steps 2 and 3 until no marked entry remains that once is strikethrough. All marked entries that are strikethrough both vertically and horizontally create cluster *O-K* for rows and cluster *F-K* for columns.
Step 5: Create a new incidence matrix A(k+1) by removing all double strikethrough entries of incidence matrix A(k). Stop algorithm when the order of matrix A(k+1) is zero, that means every entry of the matrix is double strikethrough. Otherwise, go back to step 1 by putting k=k+1. As shown in Figure 8, Figure 7 is clustered as a matrix by using CDA.

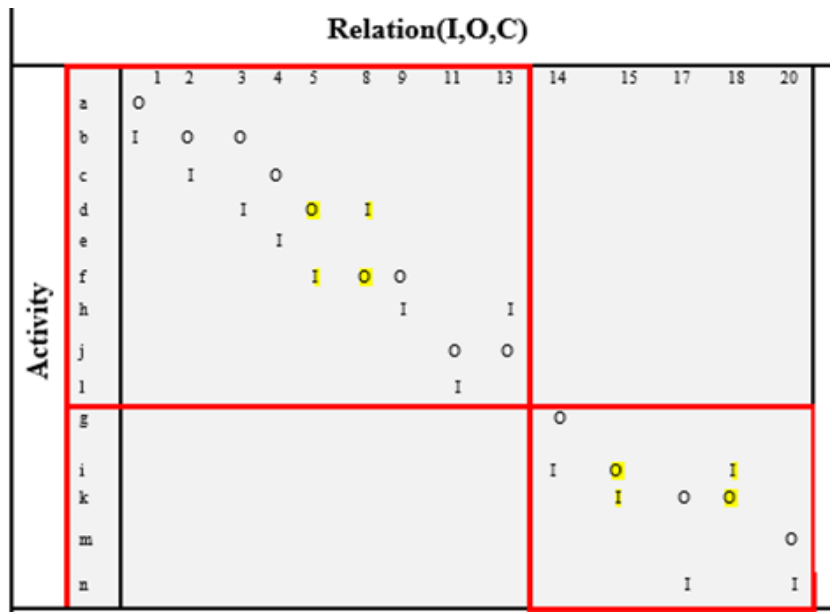

*Figure 8: Clustering ISM through CDA*

In Figure 8, each of the two blocks that have been indexed with a red line shows a group of activities. If it is required and the resources needed are provided, these two blocks can concurrently be performed. These two blocks are shown two sub-processes *(S1, S2)*. In addition to the detection of sub-processes (sub-matrix), cycles of each sub-matrix can be identified by reviewing the contents of each block. For example, it is assumed that *P* and *Q* represent two interfaces, and *K* and *L* are the representatives of two objects. *K* and *L* have interdependence if *K* impact on *L* and *L* impact on *K*. That means the input of *K* is the input of *L* and the input of *L* is the input of *K*. that can be shown as equation (1).

$$if\ (K,q) = O, (K,p) = I\ \&\ (L,q) = I, (L,p) = O \Rightarrow K \Pi L \quad (1)$$

For example, entries of *(i,13)=O, (i,15)=I, (k,13)=I,* and *(k,15)=O* display an iteration. These cycles are shown by the yellow color.

# 5- Developing a Mixed Matrix Method

The sub-processes of the process model of Figure 2 is demonstrated in Figure 9.

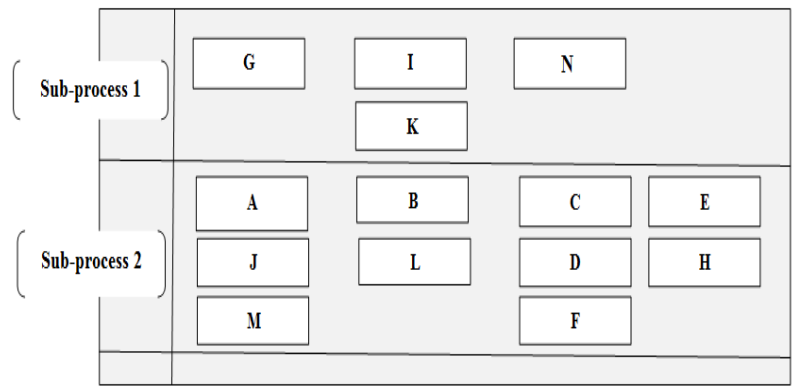

*Figure 9: Demonstration of sub-processes of Figure 2*

By drawing input/output flows of Figure 9, a process model can be created in the form of Figure 10

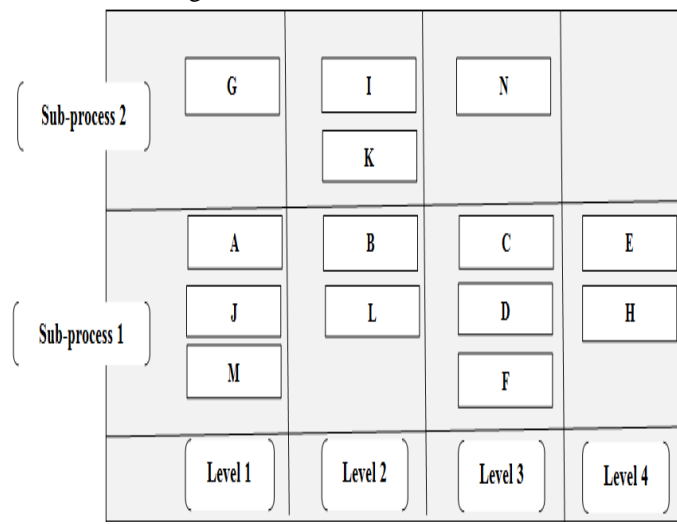

*Figure 10: **The simultaneous demonstration of activity levels and sub-processes of Figure 2***

In addition to various information that can be obtained from Figure 11, the distance of activity levels from each other also can be found. In the above figure, two rings are visible between activities *D* and *F* as well as activities *I* and *K*. This means that there is some iteration in second and third levels. When upstream activities are dependent on downstream activities (i.e., a backward dependency), there is a possibility of delay in the implementation of activities and process completion time. This possibility will be more when the distances between activities (their levels) are high. For example, the possibility of delay in the implementation of a process (or project) when some activities of levels *G* and *H* are interdependent (i.e., they form a cycle) is higher than that when one activity of *G* is dependent on one activity of *A* (i.e., they form a circulation). Anyway, the interdependence between activities is one of the major issues that designers must consider and use appropriate methods to engineer them.

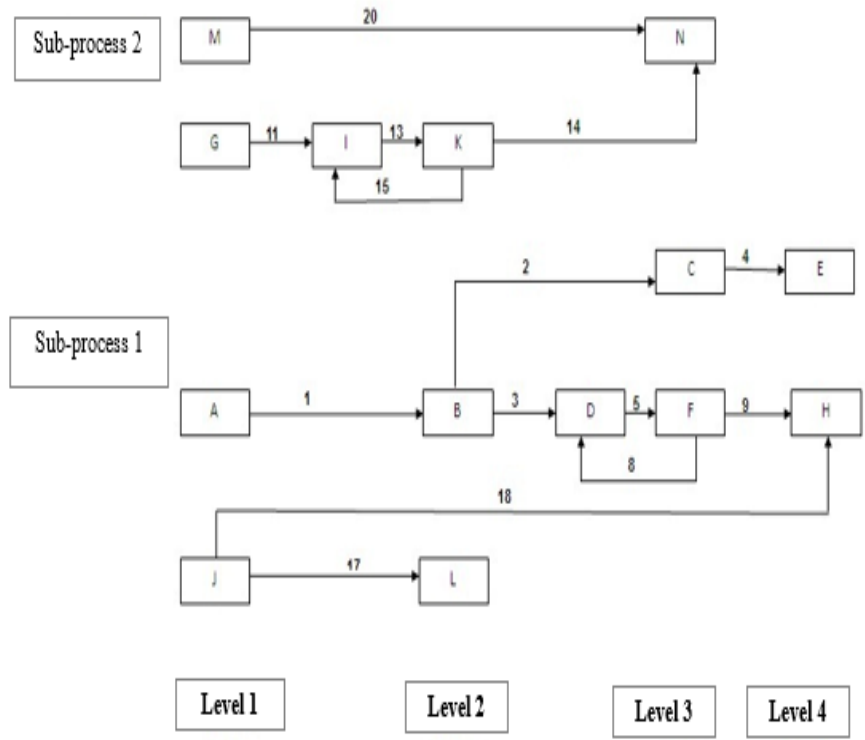

*Figure 11: **The simultaneous demonstration of activity levels and sub-processes***

## 6- Conclusion

The main idea discussed in this paper is focused on better understanding and reduction of the system's complexity. Sometimes different approaches allow analyzing various aspects of a system. Mixed application of these tools can provide a more obvious and comprehensive picture concerning the behavior of its lower-level constituents. DSM has been widely used as a powerful tool to analyze complex systems. The complexity of designing and scarcity of solution techniques as well as failure to provide the structured information about sub-processes of a system are the main limitations of this approach. Using ISM as a complementary method can overcome these limitations. If DSM is written based on the IR/FAD method, the interdependence of system elements will be displayed above the main diagonal elements. If any entry above the main diagonal is only an empty entry away, longitudinally and latitudinally, from the main diagonal, it causes iteration. Otherwise, it is considered as a cycle. The space that is created by iterations and cycles is called the space of feedback loops. Systems analysts, who want to optimize system processes and reform waste flow and create flows, and efficient patterns should have special attention to this space. Displaying simultaneously, some aspects of the system provides a better chance to engineer some flows within the system. In spite of various constraints, the above-mixed method has a clear structure and can easily be completed with other methods of data mining and mathematical modeling. Anyway, planners and managers who are interested in knowing how activities of a process should be organized appropriately and how many activity levels are in a project or how much independence exists between sub-systems can get useful information from the above-developed model. But, this model does not provide comprehensive information to those managers who are eager to know the cost of carrying out activities, the ability to predict results, and the interaction of system components. Therefore the necessity of using other complementary mathematical techniques such as discrete event simulation is strongly felt.

**REFERENCES**

Alizon, F., Moon, S. K., Shooter, S. B., & Simpson, T. W. (2007). Three dimensional design structure matrix with cross-module and cross-interface analyses. In *ASME 2007 International Design Engineering Technical Conferences and Computers and Information in Engineering Conference* (pp. 941–948). American Society of Mechanical Engineers.

Bilalis, N., Maravelakis, E., Antoniadis, A., & Moustakis, V. (2004). Mapping product innovation profile to product development activities-The I-DSM tool. In *Engineering Management Conference, 2004. Proceedings. 2004 IEEE International* (Vol. 3, pp. 1018–1022). IEEE.

Brady, T. K. (2002). Utilization of dependency structure matrix analysis to assess complex project designs. In *ASME 2002 International Design Engineering Technical Conferences and Computers and Information in Engineering Conference* (pp. 231–240). American Society of Mechanical Engineers.

Browning, T. R. (2001). Applying the design structure matrix to system decomposition and integration problems: a review and new directions. *IEEE Transactions on Engineering Management*, *48*(3), 292–306.

Chen, C.-H., Ling, S. F., & Chen, W. (2003). Project scheduling for collaborative product development using DSM. *International Journal of Project Management*, *21*(4), 291–299.

Chen, K.-M., & Liu, R.-J. (2005). Interface strategies in modular

product innovation. *Technovation*, *25*(7), 771–782.

Cho, S.-H., & Eppinger, S. D. (2005). A simulation-based process model for managing complex design projects. *IEEE Transactions on Engineering Management*, *52*(3), 316–328.

Jose, A., & Tollenaere, M. (2005). Modular and platform methods for product family design: literature analysis. *Journal of Intelligent Manufacturing*, *16*(3), 371–390.

Karniel, A., Belsky, Y., & Reich, Y. (2005). Decomposing the problem of constrained surface fitting in reverse engineering. *Computer-Aided Design*, *37*(4), 399–417.

Kreimeyer, M., Eichinger, M., & Lindemann, U. (2007). Aligning multiple domains of design processes. In *International Conference on Engineering Design* (Vol. 7, p. 162).

Kusiak, A. (1999). *Engineering design: products, processes, and systems*. Academic Press, Inc.

Kusiak, A. (2008). Interface structure matrix for analysis of products and processes. In *LCE 2008: 15th CIRP International Conference on Life Cycle Engineering: Conference Proceedings* (p. 444). CIRP.

Kusiak, A., & Wang, J. (1993). Efficient organizing of design activities. *The International Journal Of Production Research*, *31*(4), 753–769.

Marca, D. A., & McGowan, C. L. (1987). *SADT: structured analysis and design technique*. McGraw-Hill, Inc.

Mayer, R. J., Menzel, C. P., Painter, M. K., Dewitte, P. S., Blinn, T., & Perakath, B. (1995). *Information integration for concurrent engineering (IICE) IDEF3 process description capture method* *report*. KNOWLEDGE BASED SYSTEMS INC COLLEGE STATION TX.

Pektaş, Ş. T., & Pultar, M. (2006). Modelling detailed information flows in building design with the parameter-based design structure matrix. *Design Studies*, *27*(1), 99–122.

Ross, D. T. (1977). Structured analysis (SA): A language for communicating ideas. *IEEE Transactions on Software Engineering*, (1), 16–34.

Schmidt III, R., Austin, S. A., & Brown, D. (2009). Designing adaptable buildings.

Sosa, M. E., Eppinger, S. D., & Rowles, C. M. (2007). A network approach to define modularity of components in complex products. *Journal of Mechanical Design*, *129*(11), 1118–1129.

Steward, D. V. (1981). The design structure system: A method for managing the design of complex systems. *IEEE Transactions on Engineering Management*, (3), 71–74.

Suh, E. S., De Weck, O. L., & Chang, D. (2007). Flexible product platforms: framework and case study. *Research in Engineering Design*, *18*(2), 67–89.

Wiest, J. D., & Levy, F. K. (1977). A management guide to PERTICPM: with GERTIPDMIDCPM and other networks. Prentice-Hall, Inc., Englewood Cliffs.